# Phase formation and superconductivity of Fe-tube encapsulated and vacuum annealed $MgB_2$

K. P. Singh[1,2], V.P.S. Awana[1,$], Md. Shahabuddin[2], R.B. Saxena[1], Rashmi Nigam[1], M.A. Ansari[1], Anurag Gupta[1], , Himanshu Narayan[1,3], S. K. Halder[1], and H. Kishan[1]

[1]National Physical Laboratory, K.S. Krishnan Marg, New Delhi-110012, India

[2] Physics Department, Jamia Millia Islamia University, New Delhi 110025, India.

[3] Department of Physics & Electronics, National University of Lesotho Roma 180, Kingdom of Lesotho Southern Africa

We report optimization of the synthesis parameters viz. heating temperature ($T^H$), and hold time ($t^{hold}$) for vacuum ($10^{-5}$ torr) annealed and $LN_2$ (liquid nitrogen) quenched $MgB_2$ compound. These are single-phase compounds crystallizing in the hexagonal structure (space group $P_6$/mmm) at room temperature. Our XRD results indicated that for phase-pure $MgB_2$, the $T^H$ for $10^{-5}$ torr annealed and $LN_2$ quenched samples is 750 $^0$C. The right stoichiometry i.e., $MgB_2$ of the compound corresponding to $10^{-5}$ Torr and $T^H$ of 750 $^0$C is found for the hold time ($t^{hold}$) of 2.30 hours. With varying $t^{hold}$ from 1- 4 hours at fixed $T^H$ (750 $^0$C) and vacuum ($10^{-5}$ torr), the *c*-lattice parameter decreases first and later increases with $t^{hold}$ (hours) before a near saturation, while the *a*-lattice parameter first increase and later decreases beyond $t^{hold}$ of 2.30 hours. *c/a* ratio versus $t^{hold}$ plot showed an inverted bell shape curve, touching the lowest value of 1.141 which is reported value for perfect stoichiometry of $MgB_2$. The optimized stoichimetric $MgB_2$ compound exhibited superconductivity at 39.2 K with transition width of 0.6 K. In conclusion, the synthesis parameters for phase pure stoichimetric vacuum annealed $MgB_2$ compound are optimized and are compared with widely reported Ta tube encapsulated samples.

[$] Corresponding Author; e-mail: awana@mail.nplindia.ernet.in

## INTRODUCTION

Discover of superconductivity at 39 K in $MgB_2$ has attracted a lot of attention from both condensed matter experimentalists and theoreticians as well [1]. Appearance of superconductivity at 39 K seems to be just at the limit of BCS theory. In fact clear evidence of isotope effect has already given an indication that phonons play an important role in pairing mechanism in this compound [2]. The compound had yet been studied apparently rigorously in terms of its crystal structure, thermal and electrical conduction [3-5], specific heat [6, 7], isotope effect [2,8] and doping [8-10]. Based on various physical property measurements, important critical parameters of the compound viz., critical superconducting temperature ($T_c$), coherence length ($\xi$), penetration depth ($\lambda$), critical current ($J_c$) and lower/upper critical fields $H_{c1}/H_{c2}$ are yet been determined and reviewed [11, 12]. In fact the superconductivity of $MgB_2$ is much simpler in comparison to the yet mysterious HTSc (High $T_c$ superconductivity) cuprates. In particular $MgB_2$ has a simpler structure, low anisotropy and larger coherence length. Most interestingly $MgB_2$ possess higher quality grain boundaries [13], which permit excellent current transport. Higher quality grain boundaries and better superconducting critical parameters provide an edge to $MgB_2$ over widely HTSC cuprates. Higher $T_c$ of 40 K and better material properties respectively in comparison so called inter-metallic BCS type superconductors and HTSc cuprates makes $MgB_2$ a unique superconductor. Very recently an avalanche of activity had taken place in terms of high quality $MgB_2$ tapes and wires [14, 15]. Less anisotropy and high quality grain boundaries are further improved by nano-size particle doping. Nano-size various dopants prefer to stay at low angle grain boundaries and thus improve dramatically the transport properties of the compound [16, 17].

Interestingly enough though the importance of $MgB_2$ superconductor is realized by both theoreticians and experimentalists, the quality of the material in terms of its phase formation, right stoichiometery, grains size/connectivity and porosity etc. still warrants improvements. Pure phase $MgB_2$ can be synthesized in oxygen free enviorenment from 700 $^0$C to 1400 $^0$C, respectively in vacuum ($10^{-5}$ torr) to high pressure of Argon gas [18, 19]. We report here optimization of the synthesis parameters viz. heating temperature ($T^H$), and hold time ($t^{hold}$) for vacuum ($10^{-5}$ torr) annealed and $LN_2$ (liquid nitrogen) quenched Fe-tube encapsulated $MgB_2$ compound. The optimized stoichimetric $MgB_2$ compound exhibited superconductivity at 39.2 K with transition width of 0.6 K.

## EXPERIMENTAL DETAILS

Various $MgB_2$ compounds were synthesized using high quality Mg and B powder, by mixing them in stoichiometeric ratio. The mixed and ground powder, are further palletized. The pellets are than put in closed end soft iron (SS) tube. The SS tube which contains the raw $MgB_2$ pellet inside was than sealed inside a quartz tube at high vacuum of $10^{-5}$ Torr, please see Fig.1. The encapsulated raw $MgB_2$ pellet is than heated at desired heating temperatures ($T^H$) with a hold time ($T^{hold}$) and is finally quenched in liquid nitrogen ($LN_2$). X-ray diffraction (XRD) patterns were obtained at room temperature using $CuK_\alpha$ radiation. Resistivity measurements were made in the temperature range of 12 to 300 K using a four-point-probe technique on a Close Cycle refrigerator (CCR).

## RESULTS AND DISCUSSION

Figure 1 depicts the room temperature XRD (X-ray diffraction) patterns of various $MgB_2$ compounds being synthesized at a fixed $T^H$ of 750 $^0$C and different $T^{hold}$ ranging from 1 hour to 4.30 hours. All the samples crystallize in Hexagonal ($P_{6/mmm}$) structure without any noticeable impurity. Worth mentioning is the fact that for $T^{hold}$ of either less/more than 1 hour/ 5 hours with $T^H$ of 750 $^0$C the resultant material were not single phase. In fact very small impurity of MgO at 2(theta) = 37 degree is seen for the samples having $T^{hold}$ of either 1 hour or 4.30 hours. Also we found that with either higher or lower $T^H$ than 750 $^0$C, the resulting $MgB_2$ samples were not single phase. Worth mention is the fact that $T^H$ of 750 $^0$C is corresponding to vacuum annealing of $10^{-5}$ Torr. In fact, it is reported that pure phase $MgB_2$ can be synthesized in oxygen free enviorenment from 700 $^0$C to 1400 $^0$C, respectively in vacuum ($10^{-5}$ torr) to high pressure of Argon gas [18, 19]. Here in this article we focus our attention on to optimization of vacuum ($10^{-5}$ torr) annealed $MgB_2$ in its single phase region.

Lattice parameters *a* and *c* for $MgB_2$ compounds being synthesized at a fixed $T^H$ of 750 $^0$C and different $T^{hold}$ ranging from 1 hour to 4.30 hours are plotted in Fig. 3. It is seen that, the *c*-lattice parameter decreases first and later increases with $t^{hold}$ (hours) before a near saturation, while the *a*-lattice parameter first increase and later decreases beyond $t^{hold}$ of 2.30 hours. The *c/a* ratio is plotted in Fig. 4, which showed an inverted bell shape curve, touching the lowest value of 1.141. Interestingly, 1.141 (*c/a*) value correspond to $t^{hold}$ of 2.30 hours. It is reported earlier through various experimentations regarding phase formation of $MgB_2$ that *c*/*a* ratio of this compound touches the value 1.414 for perfect stoichiometry of $MgB_2$, without any deficiency of Mg [20]. This shows that we achieved perfect stoichiometry of $MgB_2$ without any Mg deficiency at $T^{hold}$ of 2.30 hours and $T^H$ of 750 $^0$C in vacuum of $10^{-5}$ torr.

Resistance (R) versus temperature (T) plots for 750 $^0$C and $10^{-5}$ Torr vacuum annealed samples with different $T^{hold}$ of 1 hour to over 3 hours are given in Fig.5. Generally speaking all the samples are metallic with superconducting transition temperature ($T_c^{R=0}$) in the range of 37 K to 39.2 K. The optimum superconductivity is achieved for $T^{hold}$ of 2.5 hours with $T_c^{R=0}$ of 39.2 K and transition width ($T_c^{onset}$ - $T_c^{R=0}$) of around 0.6 K.

As far as normal state (above $T_c^{onset}$) behavior is concerned, it is in general agreement with reported data on $MgB_2$ polycrystalline samples. The R(T) has a near constant metallic slope down to say 100 K, and later it seems to follow power law with much less positive slope. The ratio of resistance (RR), which is generally defined as $R_{300}/R_{onset}$ is close to 3.0. In literature the RR has been found up to above 6.0 in good quality $MgB_2$ samples [11, 21]. RR is found to be lower for disordered samples, for example in $MgB_{2-x}C_x$ compounds the RR comes down to say 1.5 [21]. It seems that though we achieved phase pure stoichiometric $MgB_2$ as indicated by *c*/*a* ratio of 1.413 for optimized sample, the disorder being created by possible Fe inclusion in the material can not be ignored. Generally the encapsulation done in Ta tube instead of SS-Fe tube helps in achieving the better quality of $MgB_2$. Our results indicate that SS-Fe encapsulation gives rise to Fe induced disorder in $MgB_2$ and hence is not the very right choice for attaining disorder free $MgB_2$.

In conclusion, the synthesis parameters for phase pure stoichimetric vacuum annealed $MgB_2$ compound are optimized and its superconducting properties are compared with widely reported Ta tube encapsulated samples.

## ACKNOWLEDGEMENT

Authors from the NPL appreciate the interest and advice of Professor Vikram Kumar (Director) in the present work.

**FIGURE CAPTIONS**

Fig.1 Photograph for SS-Fe tube inserted and quartz tube encapsulated $MgB_2$ raw compound.

Fig. 2 X-ray diffraction patterns for 750 $^0$C and $10^{-5}$ Torr vacuum annealed $MgB_2$ samples with different $T^{hold}$.

Fig. 3 Lattice parameters *a* and *c* versus $T^{hold}$ plot for 750 $^0$C and $10^{-5}$ Torr vacuum annealed $MgB_2$ samples.

Fig. 4 *c*/*a* versus $T^{hold}$ plot for 750 $^0$C and $10^{-5}$ Torr vacuum annealed $MgB_2$ samples.

Fig. 5 R(T) plots for 750 $^0$C and $10^{-5}$ Torr vacuum annealed $MgB_2$ samples with different $T^{hold}$.

Fig. 1

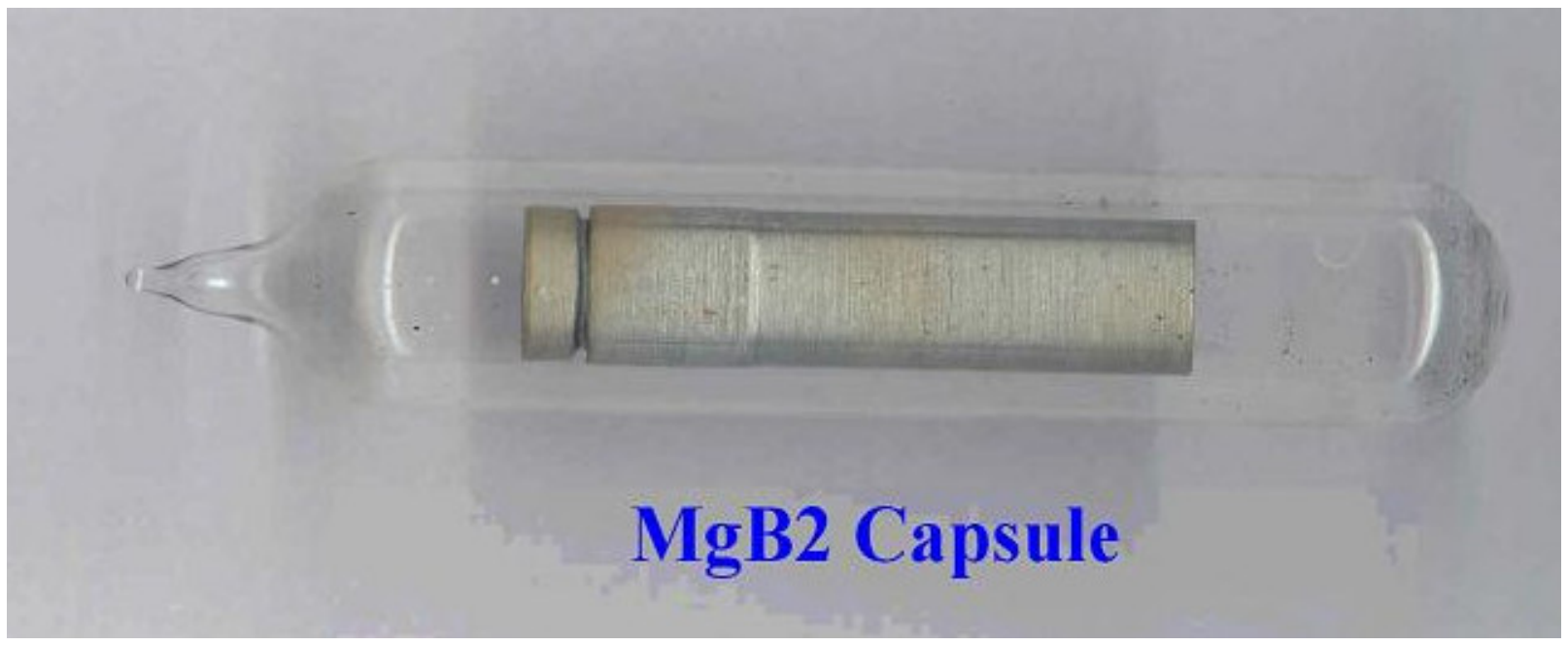
MgB2 Capsule

Fig. 2

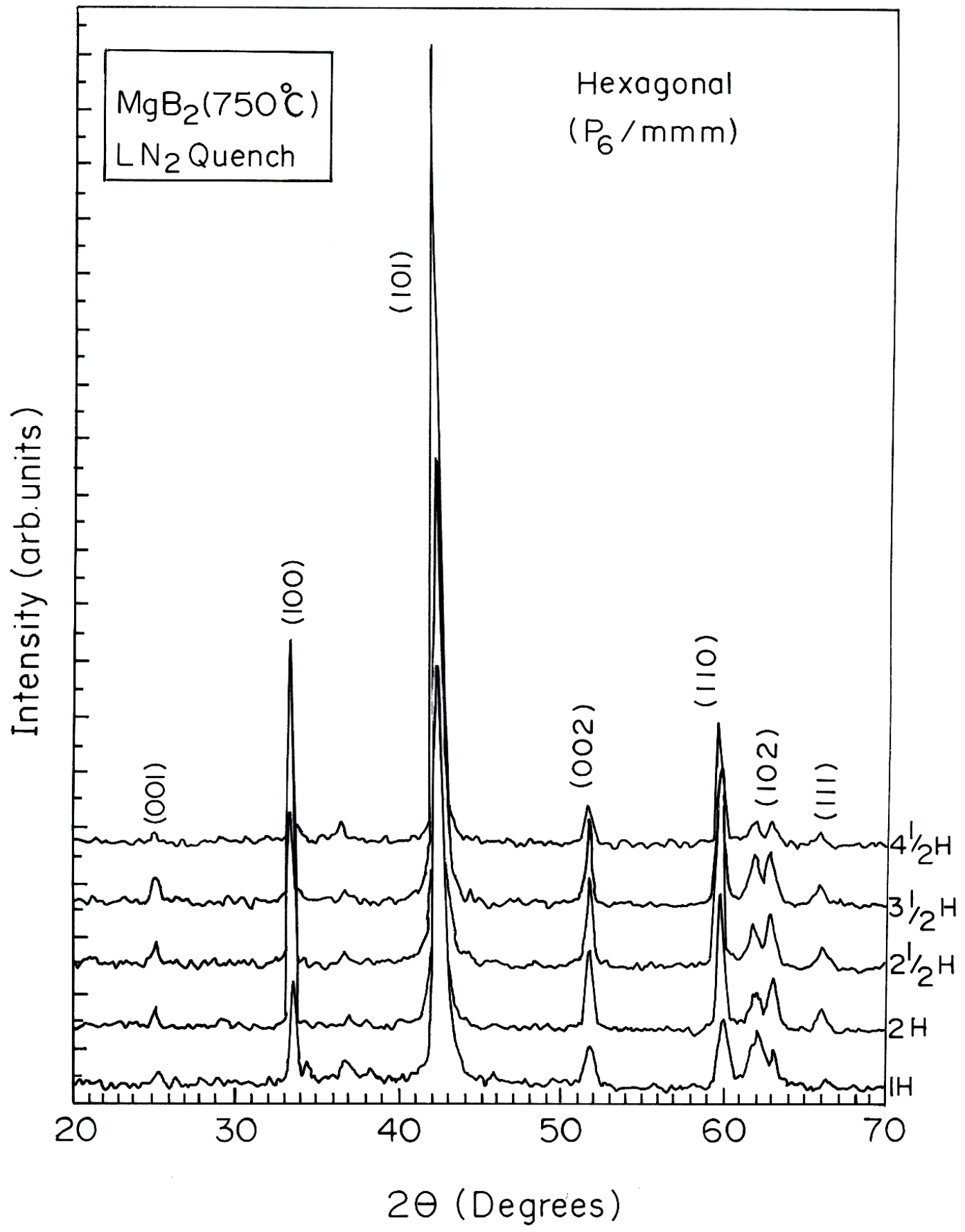

Fig.3

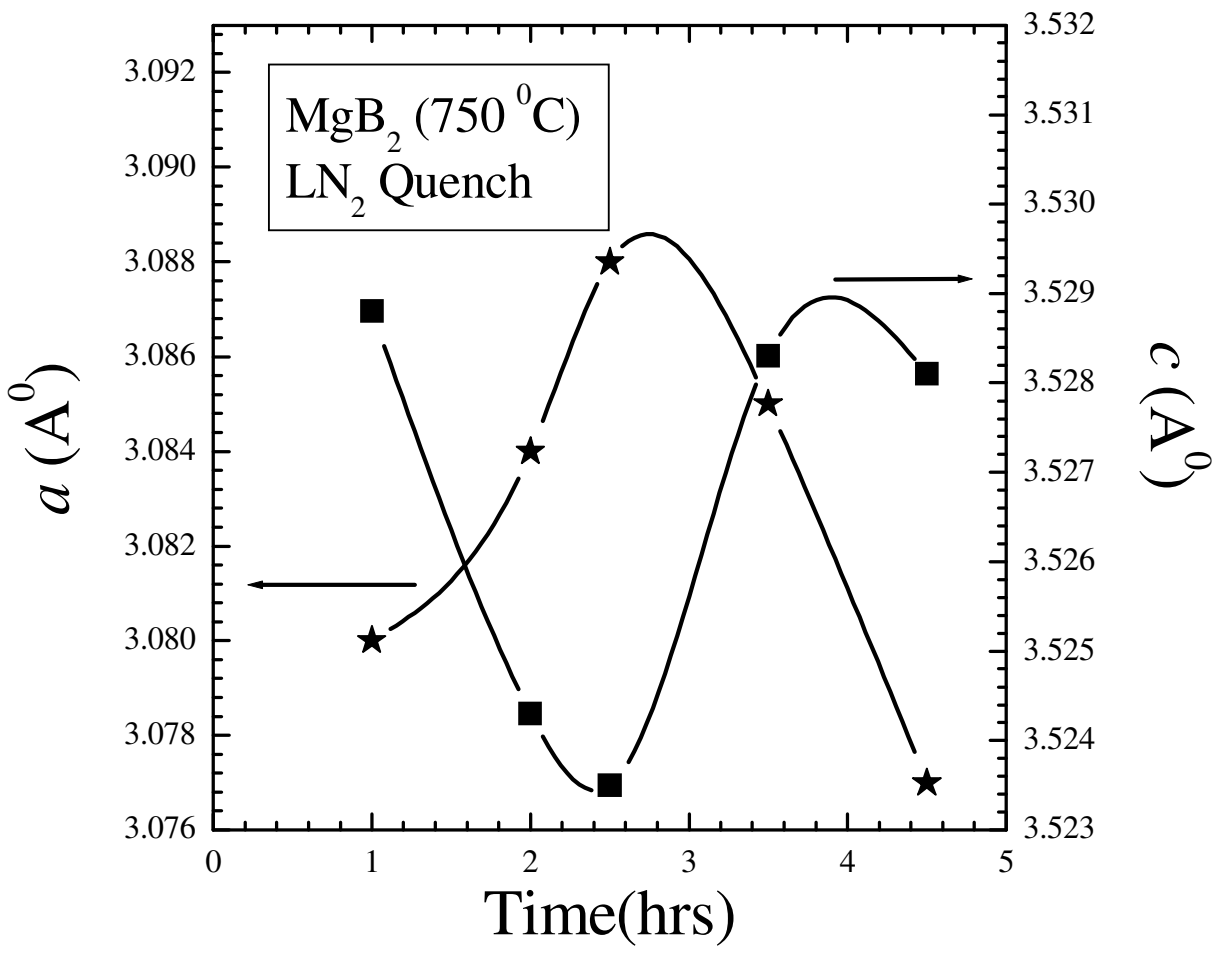

Fig.4

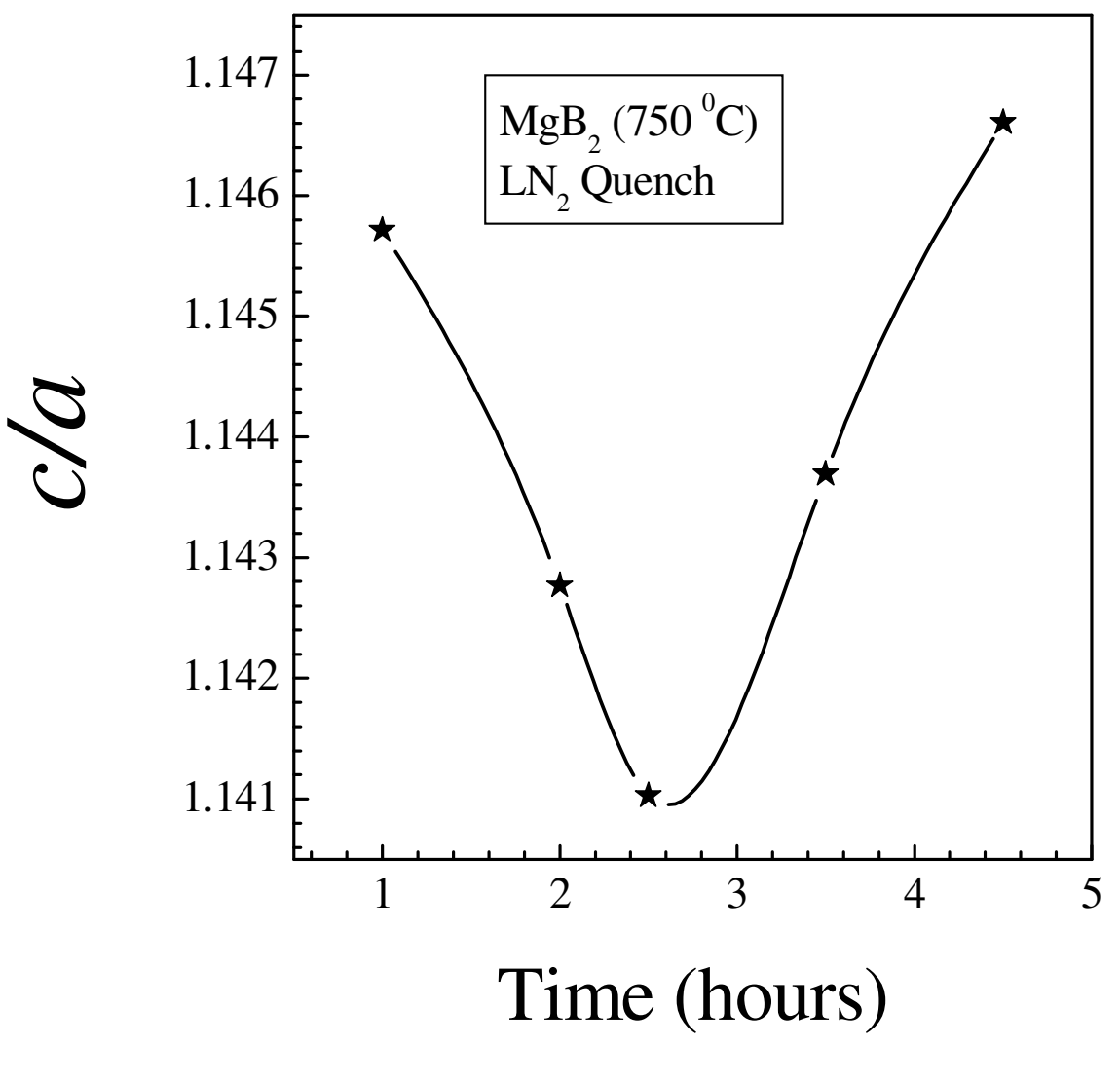

Fig. 5

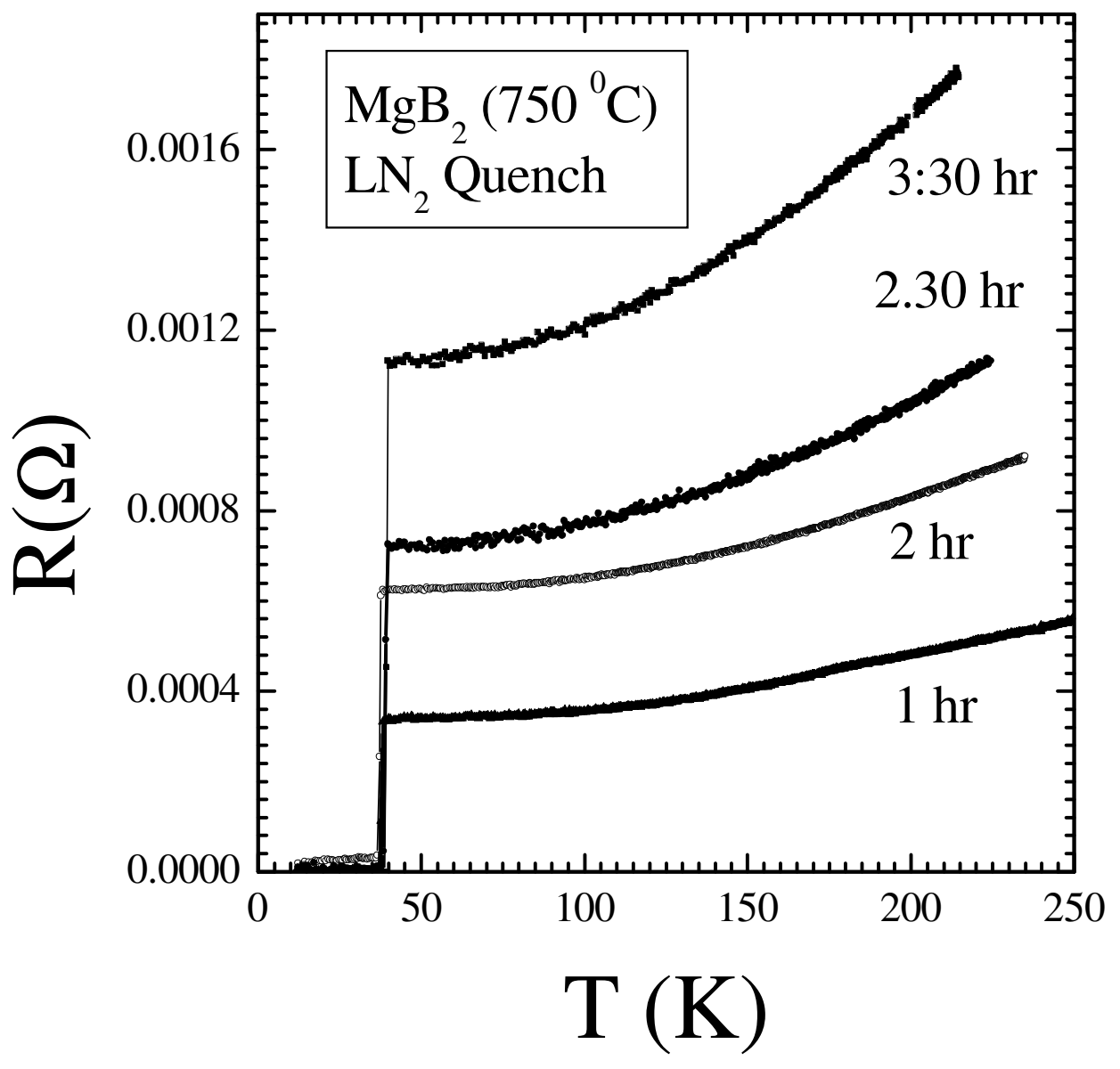